\documentclass[reprint,amsmath,amssymb,aps,prb,superscriptaddress,floatfix,nofootinbib,nobibnotes]{revtex4-2}

\pdfoutput=1
\usepackage{graphicx}
\usepackage{dcolumn}
\usepackage{bm}
\usepackage{braket}
\usepackage[mathlines]{lineno}
\usepackage{stackengine}
\usepackage{verbatim}
\usepackage{graphics}
\usepackage{hyperref}
\usepackage{xcolor}
\usepackage{appendix}
\usepackage{soul}
\usepackage{ulem} 
\usepackage{multirow}

\begin{document}
\preprint{APS/123-QED}
\title{Phase classification in the long-range Harper model using machine learning}

\author{Aamna Ahmed}\thanks{Contributed equally}
\affiliation{Department of Physics, Indian Institute of Science Education and Research, Bhopal, Madhya Pradesh 462066, India\\}
\author{Abee Nelson}\thanks{Contributed equally}
\affiliation{Department of Physics, Indian Institute of Science Education and Research, Bhopal, Madhya Pradesh 462066, India\\}
\author{Ankur Raina}
\affiliation{Department of EECS, Indian Institute of Science Education and Research, Bhopal, Madhya Pradesh 462066, India\\}
\author{Auditya Sharma}%
\affiliation{Department of Physics, Indian Institute of Science Education and Research, Bhopal, Madhya Pradesh 462066, India\\}

\date{\today}


\begin{abstract}
In this work, we map the phase diagrams of one-dimensional
quasiperiodic models using artificial neural networks. We observe that
the multi-class classifier precisely distinguishes the various phases,
namely the delocalized, multifractal, and localized phases, when
trained on the eigenstates of the long-range Aubry-Andr\'e Harper
(LRH) model. Additionally, when this trained multi-layer perceptron is
fed with the eigenstates of the Aubry-Andr\'e Harper (AAH) model, it
identifies various phases with reasonable accuracy. We
  examine the resulting phase diagrams produced using a single
  disorder realization and demonstrate that they are consistent with
  those obtained from the conventional method of fractal dimension
  analysis. Interestingly, when the neural network is trained using
the eigenstates of the AAH model, the resulting phase diagrams for the
LRH model are less exemplary than those previously obtained. Further,
we study binary classification by training the neural network on the
probability density corresponding to the delocalized and localized
eigenstates of the AAH model. We are able to pinpoint the critical
transition point by examining the metric ``accuracy" for the central
eigenstate. The effectiveness of the binary classifier in identifying
a previously unknown multifractal phase is then evaluated by applying
it to the LRH model.
\end{abstract}

\maketitle


\section{INTRODUCTION}
The application of machine learning (ML) techniques in the field of
condensed matter physics has been a subject of great interest in
recent times~\cite{doi:10.1126/science.aaa8415}. This
interdisciplinary field has rapidly gained popularity owing to the
advantages presented by machine learning~\cite{RevModPhys.91.045002,PhysRevB.96.045145,Ramprasad2017Machine}. As an actively evolving area
of research, ML techniques have been used to detect classical and
quantum phase transitions~\cite{PhysRevX.7.031038, PhysRevB.95.245134,
  PhysRevB.96.184410, PhysRevE.96.022140, PhysRevB.94.195105,
  PhysRevLett.120.176401,https://doi.org/10.48550/arxiv.1707.00663,PhysRevE.95.062122,cadez2023machine},
for the acceleration of Monte Carlo
simulations~\cite{PhysRevB.95.041101,PhysRevB.95.035105,PhysRevE.96.051301},
and for representation of states of quantum many-body
systems~\cite{doi:10.1126/science.aag2302, PhysRevX.8.011006,
  PhysRevB.99.155136,Carleo2018Constructing,Torlai2018Neural,Cai2018Approximating,Gardas2018Quantum,Carrasquilla2021How,Schmitt2020Quantum,
  Cai2018Approximating,Carleo2017Solving}. A popular subcategory of
machine learning is supervised learning~\cite{CunninghamSupervised}, where one trains algorithms
to utilize labelled training datasets in order to classify new data or
make predictions accurately. This motivates its application in
condensed matter physics for classifying various phases in strongly
correlated systems~\cite{PhysRevX.7.031038, Broecker2017}, topological
systems~\cite{PhysRevLett.118.216401,doi:10.7566/JPSJ.85.123706,
  PhysRevLett.120.066401} and quantum many-body
systems~\cite{doi:10.1126/science.aag2302, PhysRevB.94.195105}.

Disordered quantum systems~\cite{PhysRev.109.1492, RevModPhys.65.213,
  PhysRevB.35.1020, PhysRevB.104.155137, PhysRevB.106.205119,
  ahmed2023interplay} have been a fascinating topic of study for
a long time. In the 1-D Anderson
model~\cite{PhysRev.109.1492}, the tiniest of disorder is known to
localize single-particle states exponentially. In contrast, the
Aubry-Andr\`e-Harper (AAH)~\cite{aubry1980analyticity, P_G_Harper_1955} model, governed by a quasiperiodic
disorder, exhibits a
delocalization-localization transition even in one
dimension. Interestingly, instead of the exponential localization
observed in the short-ranged model, the eigenstates show algebraic
localization~\cite{PhysRevB.69.165117, PhysRevE.54.3221,
  PhysRevLett.120.110602} in the case of long-range hopping. This has
naturally led to exciting studies exploring the effects of
quasiperiodic disorder in long-range
systems~\cite{PhysRevLett.123.025301, PhysRevB.96.054202}.

The interplay of power law hopping $\left(1/r^{\sigma}\right)$ and the
quasiperiodic potential results in a rich
structure~\cite{PhysRevLett.123.025301} in the single particle
eigenstates. The self-duality of the quasiperiodic AAH model is
broken, and mobility edges are observed when the hopping is no longer
restricted to nearest neighbours only. While multifractal eigenstates
are observed to coexist with delocalized eigenstates for $\sigma<1$,
localized states exist together with delocalized eigenstates for
$\sigma>1$~\cite{PhysRevLett.123.025301}. The localization
characteristics of these single particle eigenstates have been
determined~\cite{PhysRevLett.123.025301, PhysRevB.103.075124} with the
aid of several well-known measures, such as fractal
dimension~\cite{JANSSEN1994MULTIFRACTAL,Schreiber1991Multifractal},
inverse participation
ratio~\cite{Brezini+1985+439+444,doi:10.1098/rspa.1964.0091}, level
spacing~\cite{PhysRevLett.52.1} and many more. These methods depend on
a physical understanding of the nature of the regime.

In the present work, we construct a neural network for characterizing
the phase diagram beyond established methods in single-particle
Hamiltonian models. We study the one-dimensional long-range AAH (LRH)
model, where multifractal (localized) eigenstates can coexist with
delocalized eigenstates for the long-range hopping parameter $\sigma <
1 (\sigma > 1)$. The information required for classifying the
delocalized, localized and multifractal regimes is obtained from the
eigenstates which are used as inputs to the neural network. The
trained network is first used to classify all the eigenstates of the
LRH model for various values of the hopping parameter $\sigma$. We
then consider the case of $\sigma \rightarrow \infty$, i.e., the
standard AAH model and classify its eigenstates as
delocalized/localized/multifractal. We show that the phase diagram
obtained by feeding a single disorder realization in the neural
network is in good agreement with results obtained using conventional
methods. We next train the same
network using the eigenstates of the AAH model. In this case, although
the network can identify the localized phase accurately, some
discrepancy is seen in the case of multifractal and delocalized
states. We also utilize binary classification to identify the
transition point of the AAH model by training the neural network on
its delocalized and localized eigenstates. While this network
precisely identifies the phase diagram for the AAH model, it can also
identify the multifractal phase (for which no explicit training is
given) of the LRH model, although this is achieved only in a coarse
manner.

This paper is organized as follows. Section~\ref{sec:level2} discusses
the details of the Hamiltonian model and carries an introduction to
the general setup of the neural network. In Section~\ref{sec:level3},
we discuss the network architecture used for the classification as
well as the details of the input data. In Section~\ref{sec:level4}, we
present our multi-class and binary classification results and compare
them against the results obtained through a conventional method. We
then summarize our results in Section~\ref{sec:level5}.
The utilization of the binary classifier trained on
  the data of a 1-D model, to predict the phase diagram of a $3-$D
  model is discussed in the Appendix.


\section{Model and Methods}\label{sec:level2}
\subsection{Hamiltonian}
We consider the one-dimensional long-range Harper (LRH) model given by
the Hamiltonian:
\begin{equation}
\hat{H}=-\sum_{i<j}^N\left(\frac{J}{r_{i j}^\sigma} \hat{c}_i^{\dagger} \hat{c}_j+\text { H.c. }\right)+\lambda \sum_{i=1}^N \cos \left(2 \pi \alpha i+\theta_p\right) \hat{c}_i^{\dagger} \hat{c}_i,
\label{eq1}
\end{equation}
where $\hat{c}_i^{\dagger}\left(\hat{c}_i\right)$ represents the
single particle creation (destruction) operator at site $i$. The first
term describes hopping, where $r_{i j}=\left( N/ \pi \right)
\sin\left(\pi|i-j|/N \right) $ is the geometric distance between the
sites $i$ and $j$ in a ring. The strength of the long-range hopping is
controlled by $J$, which we set to unity and the long-range parameter
in the hopping, namely $\sigma$. The second term describes the
quasiperiodic on-site energy, where the strength of the quasiperiodic
potential is $\lambda$, and the quasiperiodicity parameter $\alpha$ is
taken to be an irrational number, set as the golden mean $(\sqrt{5}-1)
/ 2$~\cite{Modugno_2009}. $\theta_p$ is an arbitrary global phase
chosen randomly from a uniform distribution in the $[0, 2\pi]$
range. The total number of sites is $N$, and periodic boundary
conditions are assumed. In the limit $\sigma \rightarrow \infty$, the
hopping is effectively nearest-neighbour, and we recover the standard
AAH model:~\cite{aubry1980analyticity, P_G_Harper_1955}
\begin{equation}
 \hat{H}=-J\sum_{i=1}^{N} (\hat{c}_i^{\dagger}\hat{c}_{i+1}+\text{H.c.})+\sum_{i=1}^{N}
 \lambda \cos(2\pi \alpha i+\theta_p)\hat{c}_i^{\dagger}\hat{c}_{i}.
 \label{eq2}
\end{equation}
It is well known~\cite{Modugno_2009} that all the energy eigenstates
are delocalized when $\lambda<2$, and all the energy eigenstates are
localized when $\lambda>2$. $\lambda=2$ is the critical point where
all the eigenstates are multifractal~\cite{RevModPhys.80.1355}. It is
known that the AAH model is self-dual~\cite{aubry1980analyticity,
  PhysRevB.28.4272}; at the critical point $\lambda=2$, the AAH model
in position space maps to itself in momentum space. When long-range
hopping is introduced ($\sigma$ is finite), the self-duality condition
is broken.

\subsection{Artificial neural network}
The artificial neural network (ANN) is inspired by the neuronal
network in the biological brain~\cite{n1,n2}. The ANN combines a
series of linear maps and non-linear
functions~\cite{Bishop2006Pattern} that are successively applied to
the input data in order to obtain the final output. A mapping defines
each layer of the network, and the dimension of a layer corresponds to
the number of neurons in it. In this work, in the case of multi-class
classification, we utilize vectors $\left\lbrace x \right\rbrace$ of
dimension $m_1$ and train a map $f(x)$ to map them to the
\textit{target set} $\left\lbrace (0,0,1),(0,1,0),(1,0,0)
\right\rbrace$. One-hot vectors represent this target set and deliver
the final output in the form of entries of the neurons of the output
layer. The network is trained on a data set called the
\textit{training set}, and its performance is gradually improved by
adjusting its parameters. In supervised learning, the training set
consists of labelled data, i.e., for each input ${x}$, the output is
already known to map to one of the outcomes of the target set. The
trained network then classifies previously unseen data from the
\textit{testing set}.

We next describe the functioning of the neural network. The `input
layer' maps the initial vectors of dimension $m_1$ to a space of
dimension $m_2$ with the help of a linear map and a non-linear
\textit{activation function}
$A$~\cite{https://doi.org/10.48550/arxiv.2109.14545}. Subsequently,
the mapping between the $n^{\text{th}}$ and the $n+1^{\text{th}}$
layer is defined as:
\begin{equation}
x^{(n)}\mapsto x^{(n+1)}=A\left( W^{(n+1,n)}x^{(n)}+B^{(n+1)}\right).
\end{equation}

Here matrix-vector multiplication is implied between $W^{(n+1,n)}$
(matrix of dimension $m_{n+1} \times m_{n}$) and $x^{(n)}$ (vector of
dimension $m_{n}$). The resulting vector $x^{(n+1)}$ is a vector of
dimension $m_{n+1}$. The elements of the matrix $W^{(n+1,n)}$ are
called the \textit{weights}, and the corresponding elements of the
vector $B^{(n+1)}$ are called the \textit{biases}. The final layer of
the ANN is referred to as the `output layer', with the number of nodes
equal to the number of expected outputs that the network is trained
for. The layers between the input and output layers are called
\textit{hidden layers}, which can range from one to several. The
method of training neural networks with multiple hidden layers is
referred to as \textit{deep learning}~\cite{GoodBengCour16}. Also, if
the number of neurons in a layer is huge, the network learns
non-universal features and unnecessary details for
classification. This can lead to \textit{overfitting} of data. Thus in
this study we employ dropout
regularisation~\cite{JMLR:v15:srivastava14a} to avoid the gradual
accumulation of neuronal weight configurations.

In the case of multi-class classification, the inputs to the network
consist of the amplitudes of an eigenstate of the Hamiltonian of
interest. For the hidden layers, we utilize the Leaky Rectified Linear Unit (ReLU) activation
function~\cite{Fukushima1975Cognitron} defined as:
\begin{equation}
  \text{Leaky ReLU:\quad } f(x_i)=
    \begin{cases}
      0.01x_i & x_i<0\\
      x_i & x_i\geq 0. 
    \end{cases}       
\end{equation}
In the output layer, we consider the Softmax activation function~\cite{Feng2019Performance}
defined as:
\begin{equation}
  \text{Softmax:\quad } f(x_i)= \frac{e^{-x_i}}{\sum_j
    e^{-x_j}}.
\end{equation}
The Softmax function estimates the output corresponding to each target
set vector, which sums up to unity. These projections can be
interpreted as the confidence of the network to assign a class to the
input data.

In the case of binary classification, since the probability densities
are considered as inputs, we utilize the ReLU activation function~\cite{agarap2019deep}:
\begin{equation}
  \text{ReLU:\quad } f(x_i)=
    \begin{cases}
      0 & x_i<0\\
      x_i & x_i\geq 0, 
    \end{cases}       
\end{equation}
in the hidden layers. In the output layer, we employ the Sigmoid
activation function~\cite{Feng2019Performance} defined as:
\begin{equation}
\text{Sigmoid:\quad } f(x_i)= \frac{1}{1+ e^{-x_i}}    
\end{equation}
to obtain a single output. There are two main advantages of the ReLU
function: (a) it is computationally efficient as it only involves a
simple comparison and (b) it introduces non-linearity in the model.
On the other hand, the Sigmoid function has a smooth and continuous
output, which makes it easier to compute gradients during
backpropagation and to optimize the model using gradient-based methods
such as the stochastic gradient descent. The
  activation functions in the hidden and output layers of both the
  neural networks are compiled in Table~\ref{tab1}.

\begin{table}[b]
\begin{tabular}{ | m{8em} | m{8em}| m{8em} | } 
  \hline
    & &  \\
  \textbf{Neural Network} & \textbf{Hidden layer activation function} & \textbf{Output layer activation function}\\ 
  \hline
    & &  \\
  Multi-class classifier & Leaky ReLU & Softmax \\
  & &  \\
  \hline
  & &  \\
  Binary classifier &  ReLU & Sigmoid \\
   & &\\
  \hline
\end{tabular}
\caption{\label{tab1}The activation functions in the
    hidden/output layer of the multi-class and binary classifiers.}
\end{table}
    
In order to correctly estimate various parameters such as the weights
and biases, the network utilizes a loss/cost
function~\cite{Goodfellow-et-al-2016} as well as an optimizer during
the training process. The cost function measures the distance between
the predicted outputs and their actual values. In general,
cross-entropy is widely used as a loss function when optimizing
classification models. For classification problems, we utilize the
cross-entropy loss function:
\begin{equation}
L_\text{CE}=\sum_{i=1}^{c_n} T_i \log S_i,
\end{equation}
where $c_n$ denotes the number of classes, $\{S_i\}$ are the
probabilities obtained from the Softmax/Sigmoid output layer, and
$\{T_i\}$ are the true values namely $0$ or $1$.

Another step involved in network training is
back-propagation~\cite{Rumelhart1986}, where the weights and biases
are adjusted in successive iterations to reduce the output of the loss
function. The updating rate of the parameters is called the
\textit{learning rate}. This is done through the \textit{gradient
  descent} optimization algorithm, which is computationally
expensive. To overcome storage issues, the training data is broken
down into small \textit{batches}, which can be easily fed to the
model. When the entire training data is fed to the model in batches,
it is called an \textit{epoch}. Thus, every batch in the training data
set can update the internal model parameters once during an epoch. In
order for the model to learn the gradient or the direction it should
take to minimize the loss function, we employ an adaptive learning
rate optimizer
\textit{Adam}~\cite{https://doi.org/10.48550/arxiv.1412.6980} which
incorporates adaptive estimates of the gradients and their
squares. The neural network evaluates the loss function on the
training dataset at the end of each epoch - this is known as the
training loss~\cite{song2016training}. An additional data set for
validation is considered for every epoch to determine whether or not
the network is fully trained, that is, if it can generalize its
knowledge to sets of previously unseen data. The neural network is
then made to evaluate the loss function on the validation dataset -
this is known as the validation loss. The training and validation
loss for a good fit should gradually reduce and converge to $0$ as the
number of epochs increases.

\section{Neural network approach}\label{sec:level3}
We perform exact diagonalization on the Hamiltonian to obtain the
single-particle eigenstates. Our goal is to be able to build an
effective neural network that can classify these eigenstates according
to their localization properties. We build and analyze two neural
networks: a multi-class and a binary classification network. While the
amplitudes of the eigenstates are utilized as inputs for the
multi-class classification network, the on-site probability densities
drawn from the eigenstates are taken to be the inputs in the case of
the binary classification network. We discuss both cases in detail
below.
  
\subsection{Multi-class classification}
We consider a system with $N=510$ sites, and obtain the
single-particle eigenstates of the LRH
model~\cite{PhysRevB.103.075124} given by Eq.~\ref{eq1}. We then
generate the training data by varying the disorder strength $\lambda$
in small steps ($0.02$) in the range of $0-5$ for the long-range
hopping parameter strengths $\sigma=0.5, 1.5$ and $3$ for several
disorder samples. As mentioned earlier, the multifractal (localized)
eigenstates coexist with delocalized eigenstates for $\sigma<1\left(
\sigma>1 \right)$ in the LRH model; the chosen parameters help us to
obtain data corresponding to all the three classes, i.e., delocalized,
multifractal and localized.
\begin{figure}[b]
\centering
\stackunder{}{\includegraphics[width=7.5cm]{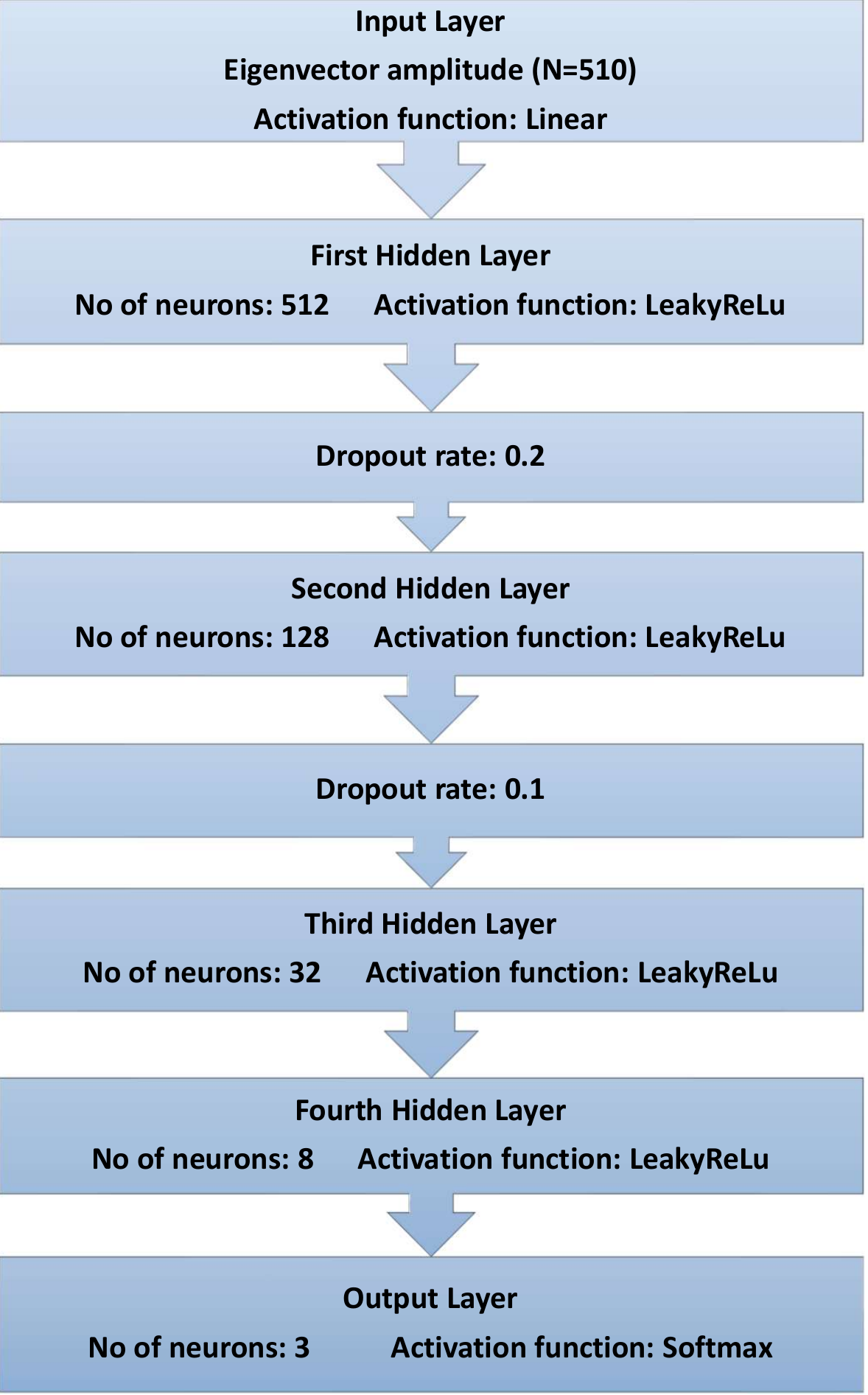}}
\caption{\label{fig1}A schematic diagram of the neural network architecture for multi-class classification of eigenstates. Here the input layer is equal to system size $N$, and the output layer has $3$ neurons, which gives the confidence of the state being delocalized, multifractal and localized. Nonlinearities are introduced by the Leaky rectified linear unit (Leaky ReLU). Dropout
is included to increase classification accuracy.}
\end{figure}


\begin{figure}
\centering
\stackunder{\hspace{-3.8cm}(a)}{\includegraphics[width=4.2cm]{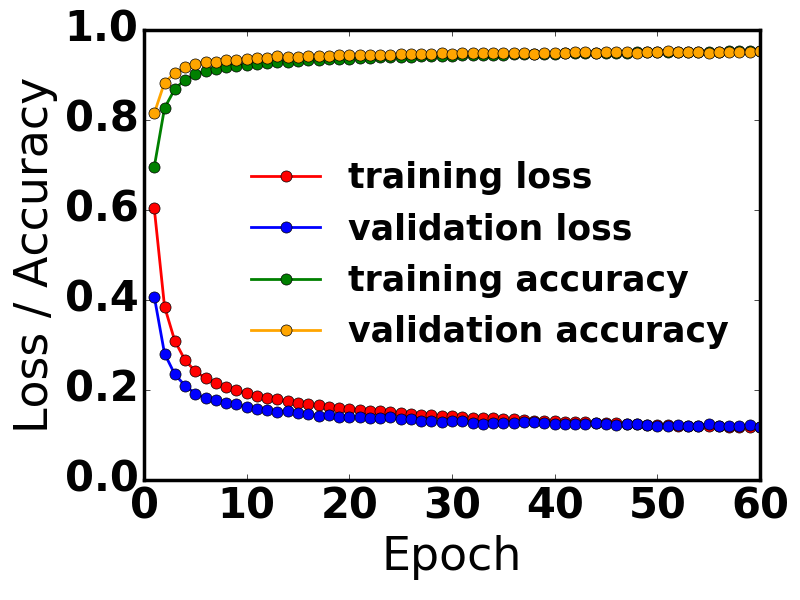}}
\stackunder{\hspace{-3.8cm}(b)}{\includegraphics[width=4.2cm]{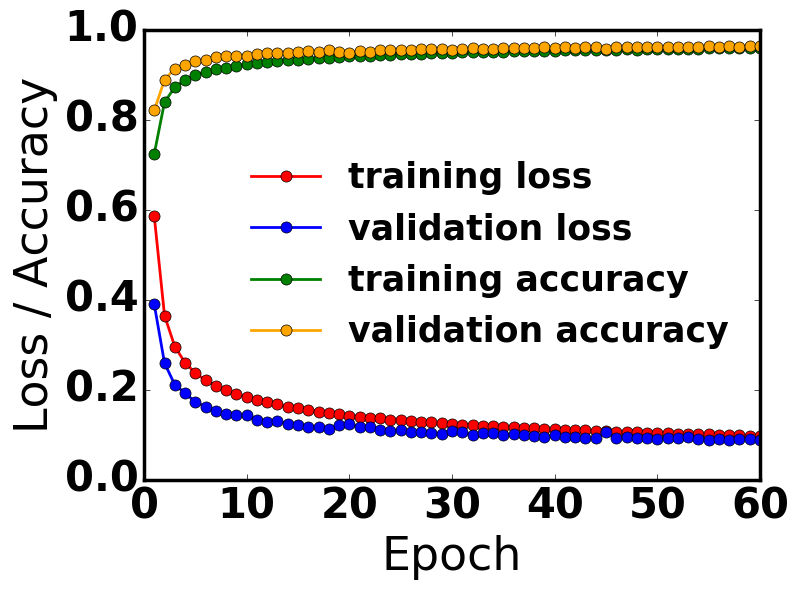}}
\caption{\label{fig2}The training loss and training accuracy along with the validation loss and validation accuracy versus the number of epochs for the neural network trained using eigenstates of the (a)~LRH model and (b)~AAH model. The stabilization in loss occurs around $50$ epochs and beyond, indicating that the ANN is trained.}
\end{figure}

In order to label the eigenstates required for network training, we calculate a well-known quantity called fractal dimension $D_q$~\cite{RevModPhys.80.1355, PhysRevLett.122.106603, PhysRevLett.123.180601} by coarse-graining the system into boxes of length $l$. Given a normalized wave function $\left|\psi_{k}\right\rangle=$ $\sum_{i=1}^N \psi_{k}(i)|i\rangle$ defined over a lattice of size $N$, we divide the lattice into $N / l$ segments of length $l$~\cite{Wang2016}. The fractal dimension is then defined as:  
\begin{equation}
D_{q}=\frac{1}{q-1}\frac{\log\sum_{p=1}^{N / l}\left[\sum_{i=(p-1) l+1}^{p l}|\psi_{k}(i)|^{2q}\right]}{\log[l/N]}.
\label{eq6}
\end{equation}

The fractal dimension in the limit $N\rightarrow \infty$ is given
by~\cite{RevModPhys.80.1355}:
\begin{equation}
D_q^{\infty}=\text{lim}_{N\rightarrow \infty}D_q.
\end{equation}

For a perfectly delocalized state, $D_q^{\infty}=1$ while for a
localized state, $D_q^{\infty}$ is vanishing, for all
$q>0$. For intermediate cases, $0<D_{q}^{\infty}<1$, which is a sign
that the state is multifractal. 

Here for each eigenstate, we calculate the fractal dimension $D_2$ and label it as follows: 
\begin{align}
\nonumber D_2 <0.2 & \qquad \text{Localized,}\\
\nonumber 0.2 \leq D_2 \leq 0.8 & \qquad \text{Multifractal,}\\
D_2 > 0.8 & \qquad \text{Delocalized}.   
\label{eq3}
\end{align}


\begin{figure}[b]
\centering
\stackunder{}{\includegraphics[width=9cm]{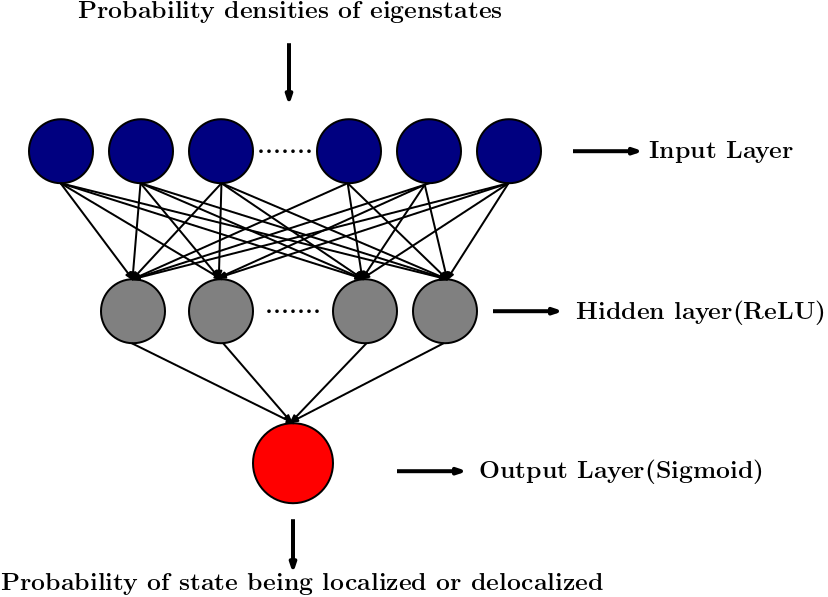}}
\caption{\label{fig3} Neural network architecture for binary classification of delocalized and localized eigenstates. Here nonlinearities are introduced by the rectified linear unit (ReLU).}
\end{figure}

\begin{figure*}
\centering
\stackunder{\hspace{-5cm}(a)}{\includegraphics[width=5.6cm]{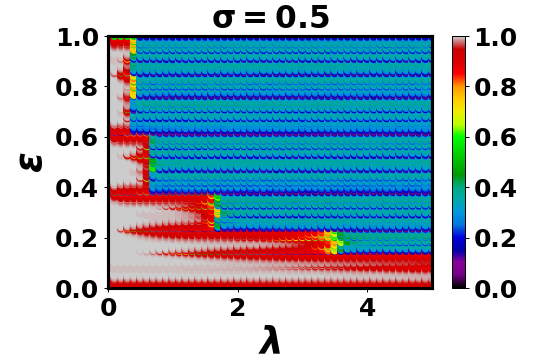}}
\stackunder{\hspace{-5cm}(b)}{\includegraphics[width=5.6cm]{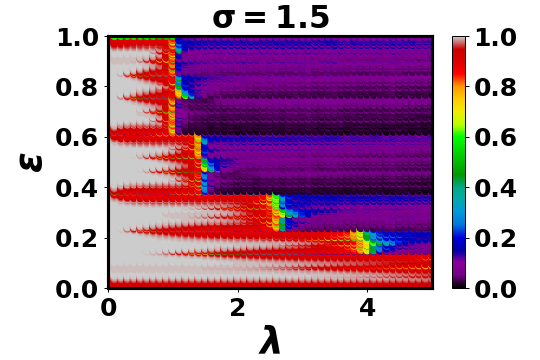}}
\stackunder{\hspace{-5cm}(c)}{\includegraphics[width=5.6cm]{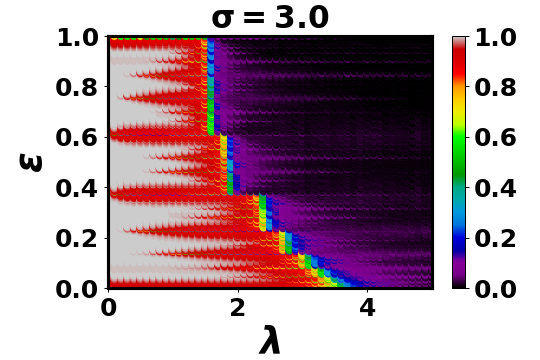}}
\vspace{-0.5cm}

\stackunder{\hspace{-5cm}(d)}{\includegraphics[width=5.6cm]{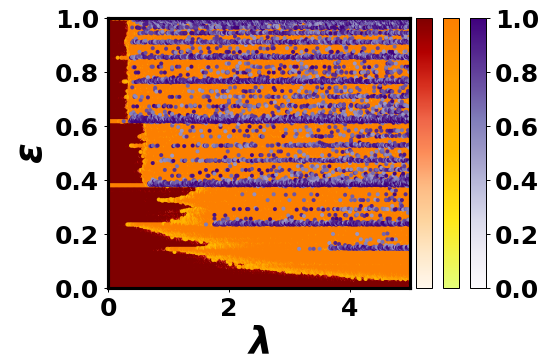}}
\stackunder{\hspace{-5cm}(e)}{\includegraphics[width=5.6cm]{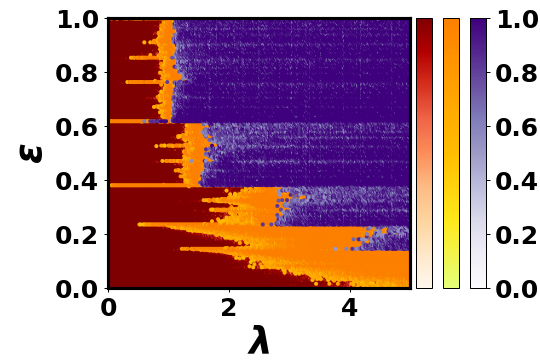}}
\stackunder{\hspace{-5cm}(f)}{\includegraphics[width=5.6cm]{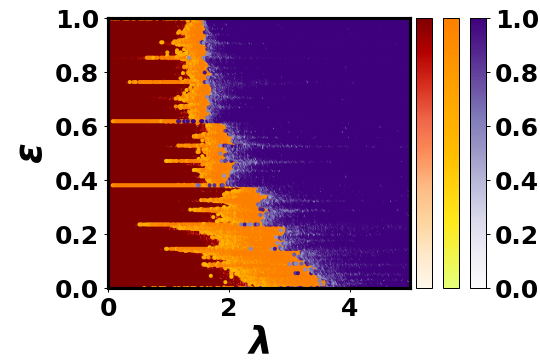}}
\vspace{-0.5cm}

\stackunder{\hspace{-5cm}(g)}{\includegraphics[width=5.6cm]{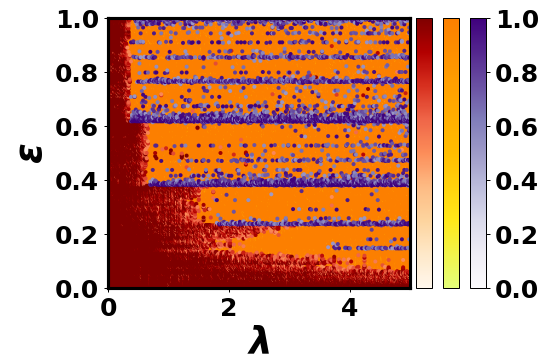}}
\stackunder{\hspace{-5cm}(h)}{\includegraphics[width=5.6cm]{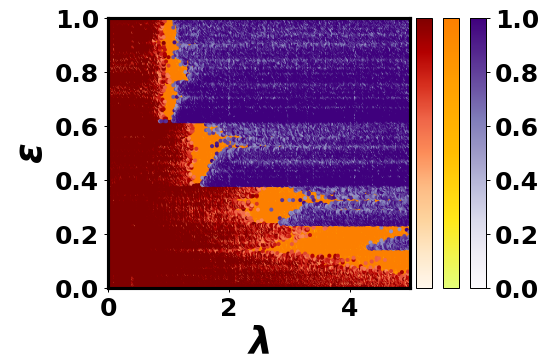}}
\stackunder{\hspace{-5cm}(i)}{\includegraphics[width=5.6cm]{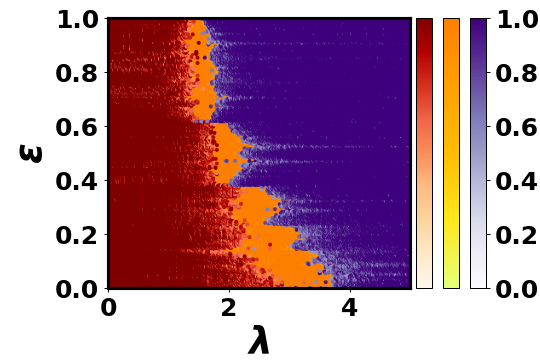}}
\caption{\label{fig4}Fractal dimension $D_2$ (whose value is
  represented by a colour according to the code shown) as a function
  of disorder strength $\lambda$ and for increasing fractional
  eigenstate index $i/N$ starting from the ground state for the
  long-range hopping parameter $\sigma$ equal to (a)~$0.5$, (b)$1.5$,
  and (c)~$3.0$. Here averaging has been performed over $100$ disorder
  realizations. Classification of the single disorder realization of
  the LRH model using the network trained on the eigenstates of the
  LRH model for hopping parameter $\sigma$ equal to (d)~$0.5$,
  (e)$1.5$, and (f)~$3.0$. Classification of the single disorder
  realization of the LRH model using the network trained on the
  eigenstates of the AAH Model for hopping parameter $\sigma$
  (g)~$0.5$, (h)$1.5$, and (i)~$3.0$. The colour of each point
  $(\sigma,\lambda)$ represents the confidence of the network for the
  delocalized (red), multifractal (orange), and localized (purple)
  phases. Here system size is $N=510$ in all cases. }
\end{figure*} 


We first generate a training data set comprising of $300000$
eigenstates of the LRH model belonging to each of the three classes,
by considering various values of $\sigma$, $\theta$, and $\lambda$. A
schematic flowchart to describe the complete architecture of the
neural network performing multi-class classification is shown in
Fig.~\ref{fig1}. The network comprises of an input layer, several
hidden layers and an output layer coupled to Linear, Leaky ReLU and
Softmax activation functions, respectively. We have also added dropout
layers to the model to avoid overfitting. The cost function is
cross-entropy, and the neural network weights are optimized using the
Adam optimizer. We consider batch sizes of $500$ samples, which add up
to $900000$ samples per epoch, with an $80\%-20\%$ split between
training and validation. We observe that
fixing the number of training epochs to $60$ allows us to obtain an
accuracy of $\approx 95\%$ as shown in Fig.~\ref{fig2}(a). Here
accuracy is defined as the ratio of eigenstates correctly classified
to the total number of eigenstates. Once the network is trained, the
eigenvectors from each point in the phase space of
$\left(\lambda,\sigma\right)$ are fed to the network. We obtain three
values each of which lies between $0$ and $1$ and correspond to the
neural network's confidence of classifying the given input in each
phase/class.

We next generate another training data set comprising of eigenstates
of the AAH model given by Eq.~\ref{eq2}. Once again the training set
consists of $300000$ eigenstates belonging to each class classified as
localized, multifractal and delocalized using $D_2$ (see
Eq.~\ref{eq3}). The network architecture is the same as before (see
Fig.~\ref{fig1}). The neural network weights are tuned using an Adam
optimizer, and the cost function is cross-entropy as before. Here
again $60$ training epochs have been considered, and the batch size is
$500$. The $80\%-20\%$ split between training and validation is
followed once again. As demonstrated in Fig. ~\ref{fig2}(b), the
network is trained to an accuracy of $\approx 95\%$.

\begin{figure*}
\centering
\stackunder{\hspace{-5cm}(a)}{\includegraphics[width=5.6cm]{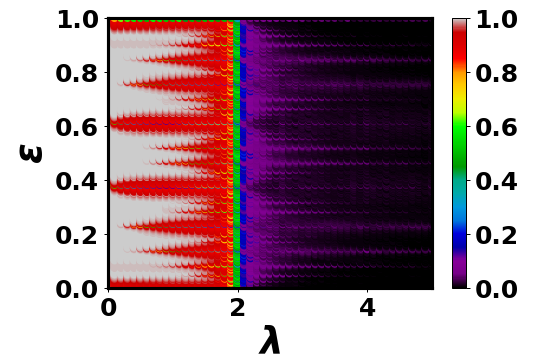}}
\stackunder{\hspace{-5cm}(b)}{\includegraphics[width=5.6cm]{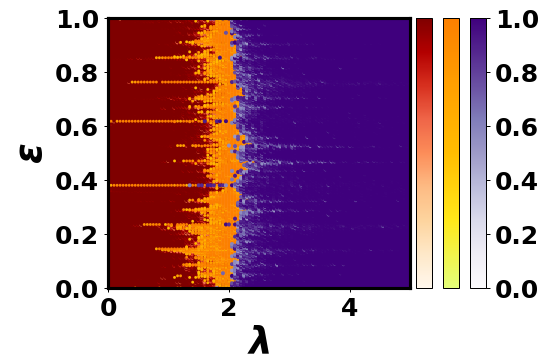}}
\stackunder{\hspace{-5cm}(c)}{\includegraphics[width=5.6cm]{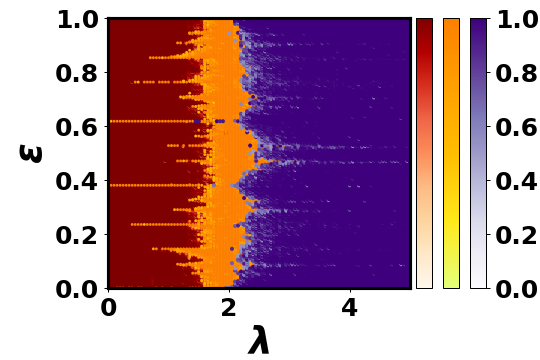}}
\caption{\label{fig5}Fractal dimension $D_2$ (whose value is
  represented by a colour according to the code shown) as a function
  of disorder strength $\lambda$ and increasing fractional eigenstate
  index $i/N$ starting from the ground state. Here averaging has been
  performed over $100$ disorder realizations. Classification of the
  single disorder realization of the eigenstates using the network
  trained on the eigenstates of the (b)~LRH Model and (c)~AAH
  model. The color of each point indicates the confidence for the
  delocalized (red), multifractal (orange), and localized (purple)
  phases. Here system size $N=510$ in all cases.}
\end{figure*} 

\subsection{Binary classification}
For binary classification, we train the network using the on-site
probability densities (PDs) of the delocalized and localized
eigenstates of the AAH model. As mentioned earlier, all the
eigenstates are delocalized (localized) below (above) a critical
disorder strength $\lambda = 2$, at which the eigenstates show
multifractal behaviour. For a system size of $N=510$, we consider
$100$ disorder realizations ($\theta_i$) generating $51000$ samples
separately for the delocalized and localized classes by choosing
disorder strengths $\lambda=0.5$ and $4$, respectively. We construct a
neural network which consists of an input layer of size $N$ and
utilizes on-site probability densities of the eigenstates as
inputs. This is followed by a single hidden layer with the number of
neurons equal to the greatest integer less than $\left( N \times (2/3)
\right) +1$. The neurons are coupled to the ReLU activation function
and an output layer coupled to the Sigmoid activation function. The
single output obtained is a number that lies between $0$ and $1$ - a
value close to $0$ indicates that the state is delocalized while a
value close to $1$ indicates that it is localized. The scheme of the
neural network architecture is shown in Fig.~\ref{fig3}. Here the cost
function is binary cross-entropy, and the optimizer is Adam. The batch
size is $300$ samples, and $5$ training epochs are considered - we
find that this is sufficient to obtain an accuracy of $\approx 100 \%
$, when the training and validation data set is split into an $80:20$
ratio. We use the metric ``accuracy'' to identify the phase
transitions. By training on two phases only, we also examine the
ability of the neural network to identify a previously unknown phase.

\section{Results}\label{sec:level4}
In this section, we perform multi-class and binary classification
using the eigenvectors and probability densities, respectively. The
results obtained using the neural network approach are compared to the
ones acquired from the conventional multifractal analysis of $D_q$.

\subsection{Multi-class classification}
We begin our analysis by calculating the multifractal dimension $D_q$
as a function of disorder strength $\lambda$ for all the single
particle eigenstates of the LRH Hamiltonian for $\sigma=0.5, 1.5$ and
$3.0$ as shown in Fig.~\ref{fig4}(a)--\ref{fig4}(c) respectively. In
the case of $\sigma=0.5$, with an increase in disorder strength, the
fraction of delocalized eigenstates decreases, and multifractal states
are observed to appear in blocks, as shown in Fig.~\ref{fig4}(a). Thus
a delocalized-to-multifractal (DM) edge is observed in the eigenstate
spectrum, which changes in a step-like fashion with increasing
disorder strength $\lambda$. In the case of $\sigma=1.5$ (see
Fig.~\ref{fig4}(b)) and $\sigma=3.0$ (see Fig.~\ref{fig4}(c)), we
observe that the delocalized eigenstates are separated from the
localized eigenstates, with a delocalized-to-localized (DL) edge which
changes in a step-like fashion as the number of delocalized states
decreases with increasing disorder strength $\lambda$. We also observe
the presence of multifractal states in the vicinity of the DM / DL
edges.

We next employ the multi-class classification algorithm for a single
disorder realization and compare its performance with the multifractal
analysis. First, the network (see Fig.~\ref{fig1}) is trained using the
eigenstates of the LRH model generated over multiple disorder
realizations for various values of $(\sigma,\lambda)$. We assign the
class to the vectors in the training data set with the help of $D_2$
using Eq.~\ref{eq3}. The phase diagram obtained as a function of
disorder strength $\lambda$ using the neural network for a single
disorder realization is shown in Fig.~\ref{fig4}(d)--\ref{fig4}(f)
corresponding to $\sigma=0.5, 1.5$ and $3$ respectively. The states
are classified as delocalized, multifractal and localized with
confidence $p_1, p_2$ and $p_3$, respectively. In all figures, we have
plotted the $max\left(p_1,p_2,p_3\right)$ where the red, orange and
purple colour codes are used to represent $p_1$, $p_2$, and $p_3$
respectively. Comparing Figs.~\ref{fig4}(a)--\ref{fig4}(c) with
Figs.~\ref{fig4}(d)--\ref{fig4}(f), we observe that the shape and
location of the transitions agree very well at all values of $\sigma$.

Next, we compare the neural network-based transition characterization
when the same network (see Fig.~\ref{fig1}) is trained using the
eigenstates of the AAH model. Although the existence of a critical
disorder strength at which all eigenstates are multifractal and
separate the delocalized and localized eigenstates is well-known, we
still assign the class to the vectors in the training data set with
the help of $D_2$ using Eq.~\ref{eq3}. Once the network is trained, we
utilize it to obtain the phase diagrams of the LRH model as a function
of disorder strength $\lambda$ using the neural network for a single
disorder realization as shown in Fig.~\ref{fig4}(g)--\ref{fig4}(i)
corresponding to $\sigma=0.5, 1.5$ and $3$ respectively. While the
network can precisely predict the location of the transitions/steps, the
distinction between the delocalized and multifractal states is not
very sharp. Nevertheless, we still observe the step-like features and
the distinct phases.

We next compute the multifractal dimension $D_q$ as a function of
disorder strength $\lambda$ for all the single particle eigenstates of
the AAH model (see Fig.~\ref{fig5}(a)). All the eigenstates are
delocalized for $\lambda < 2$ while all eigenstates are localized for
$\lambda >2$. At $\lambda=2$, the eigenstates are
multifractal~\cite{RevModPhys.80.1355,Modugno_2009}. We obtain the
phase diagram of the AAH model using the neural network trained on the
eigenstates of the LRH model (see Fig.~\ref{fig5}(b)) and the
eigenstates of the AAH model (see Fig.~\ref{fig5}(c)). The neural
network can accurately predict the phase diagram using a single
disorder realization in both cases. We also checked
  that a network trained with a higher-order fractal dimension like
  $D_6$ is also able to faithfully reproduce the phase diagram as
  shown in the Appendix.

\begin{figure}[b]
\centering
\stackunder{}{\includegraphics[width=7cm]{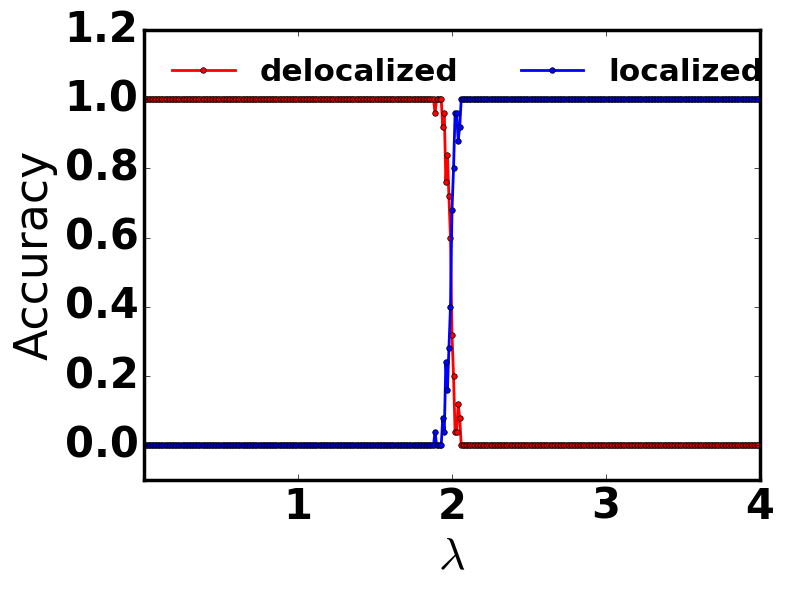}}
\caption{\label{fig6}Neural Network classification accuracy with
  increasing disorder strength $\lambda$ for the central eigenstate of
  the AAH model for system size $N=510$ averaged over $25$ disorder
  realizations.}
\end{figure}
\begin{figure}
\centering
\stackunder{\hspace{-5.8cm}(a)}{\includegraphics[width=6cm]{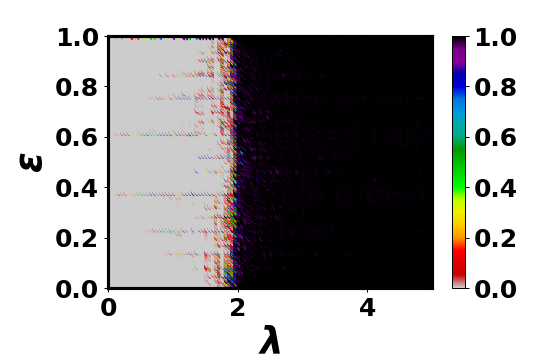}}\vspace{-0.5cm}
\stackunder{\hspace{-5.8cm}(b)}{\includegraphics[width=5.5cm]{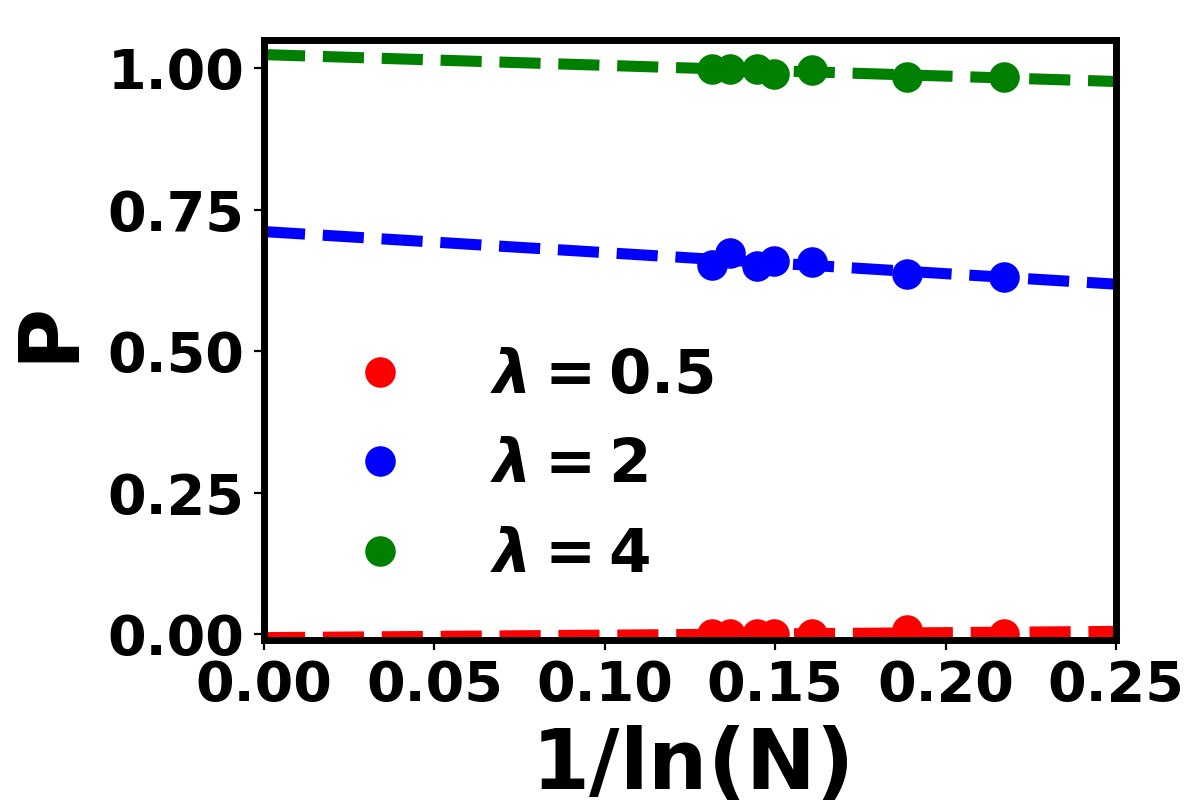}}
\caption{\label{fig7} (a)~Classification of the eigenstates for a
  single disorder realization of the AAH model as a function of
  disorder strength $\lambda$ and for increasing fractional eigenstate
  index $i/N$ starting from the ground state. Here the network is
  trained on the delocalized ($\lambda=0.5$) and localized eigenstates
  ($\lambda=4$) of the AAH model. The probability of the state being
  localized is given by the single output $P$, whose value is
  represented by a colour according to the code shown. The
  corresponding plot using fractal dimension $D_q$ is shown in
  Fig.~\ref{fig5}(a). Here system size is
  $N=510$. (b)~The probability $P$ of the state being
    localized vs $\frac{1}{\ln(N)}$ in the delocalized phase
    $\lambda=0.5$, at the critical point $\lambda=2$ and in the
    localized phase with $\lambda=4$. Here, system sizes range from $N
    = 100$ to $N = 2000$.}
\end{figure}

\begin{figure*}
\centering
\stackunder{\hspace{-5cm}(a)}{\includegraphics[width=5.6cm]{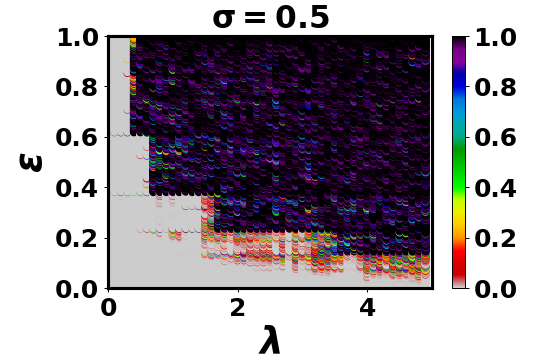}}
\stackunder{\hspace{-5cm}(b)}{\includegraphics[width=5.6cm]{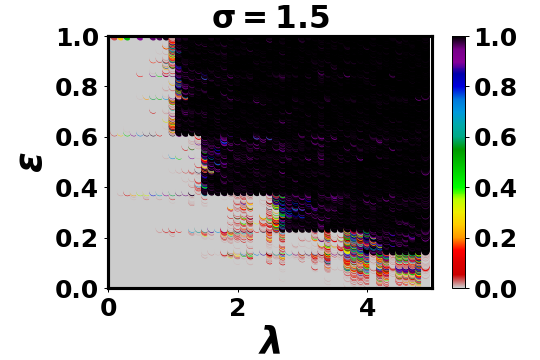}}
\stackunder{\hspace{-5cm}(c)}{\includegraphics[width=5.6cm]{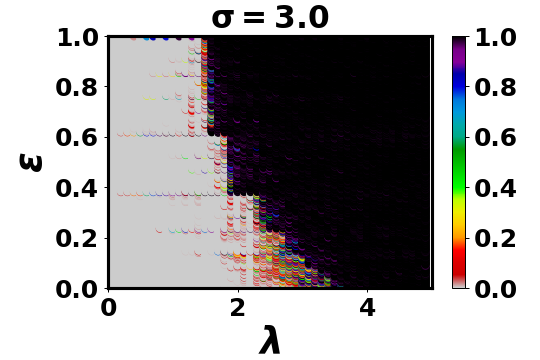}}
\caption{\label{fig8} Classification of the eigenstates for a single
  disorder realization of the LRH model as a function of disorder
  strength $\lambda$ and with increasing fractional eigenstate index
  $i/N$ starting from the ground state for the long-range hopping
  parameter $\sigma$ equal to (a)~$0.5$, (b)~$1.5$ and (c)~$3.0$. Here
  the network is trained on the delocalized ($\lambda=0.5$) and
  localized eigenstates ($\lambda=4$) of the AAH model. The
  probability of the state being localized is given by the single
  output $P$, whose value is represented by a colour according to the
  code shown. The corresponding plot using fractal dimension $D_q$ is
  shown in Fig.~\ref{fig4}(a)--\ref{fig4}(c). Here system size is
  $N=510$ in all cases.}
\end{figure*}

\subsection{Binary classification}
In this subsection, we discuss the results obtained with a binary
classifier using the neural network shown in Fig.~\ref{fig3}. The
training data set consists of the probability densities (PDs)
corresponding to the eigenstates of the two classes, i.e., delocalized
($\lambda=0.5$) and localized ($\lambda=4$). Since we have a single
neuron in the output layer, the output represents the probability $P$
of the state being localized, implying that $1-P$ is the probability
of the state being delocalized. We may infer that intermediate values
of $P$ indicate that the state exhibits multifractal nature.

The information about the transition from the delocalized to localized
regimes is incorporated in the properties of the eigenstates. We feed
the central eigenstate (corresponding to the energy $E_{N/2}$ of the
spectrum) to the trained network to determine the transition point. In
Fig.~\ref{fig6}, we have plotted the classification prediction as a
function of disorder strength $\lambda$. The transition point is
revealed to be at $\lambda=2$ as the network learns the difference
between the localized and delocalized eigenstates. The transition
point obtained agrees with the one shown using multifractal dimension
$D_2$ in Fig.~\ref{fig5}(a).
 
Next, we test this trained neural network, by inputting the
probability densities drawn from all the eigenstates of the AAH model,
as $\lambda$ is varied across the transition. The neural network gives
a single output $P$, which signifies the confidence of the network to
classify the state as localized, indicating that $1-P$ is the
probability of classifying the state as delocalized. For a single
disorder realization, the phase diagram obtained from the neural
network shown in Fig.~\ref{fig7}(a) is consistent with the one
obtained using multifractal dimension $D_2$ as shown in
Fig.~\ref{fig5}(a). We also note that at the transition point where
the multifractal states exist, the prediction $P$ lies roughly midway
between $1$ and $0$, consistent with theoretical results.
In Fig.~\ref{fig7}(b), we numerically study the
  system-size dependence~\cite{PhysRevB.106.205119} of the probability
  $P$ of the eigenstates being localized obtained using the binary
  classifier. While in the delocalized phase at $\lambda=0.5$, $P$
  remains close to $0$, at the multifractal point $\lambda=2$, it lies
  between $0$ and $1$. In the localized phase with $\lambda=4$, $P$
  approaches unity. This analysis is consistent with the ones obtained
  using conventional methods and can be utilized in classifying phases
  robustly against increasing system sizes.

Next, we implement the binary classification algorithm on the LRH
model to investigate how well our neural network can identify a
previously unknown phase. Since the network is trained exclusively on
the delocalized and localized phases, it is unfamiliar with the
multifractal regime. In Fig.~\ref{fig8}(a)--\ref{fig8}(c), we have
plotted the phase diagram by feeding the eigenstates of a single
disorder realization of the LRH model to the binary classifier
corresponding to $\sigma=0.5, 1.5$ and $3.0$ respectively. We observe
that our network can indicate the presence of a new phase
(multifractal phase). In Fig.~\ref{fig8}(a), for $\sigma=0.5$, while
the network shows confusion in the multifractal phase, the step-like
features that distinctly separate the delocalized states can still be
observed. In Fig.~\ref{fig8}(b)--(c), for $\sigma=1.5$ and $3.0$, we
observe that the delocalized and localized states are classified and
separated by step-like edges along which multifractal states are
observed with $0<P<1$. In other words, while the network identifies
the delocalized and localized phases, in the case of the multifractal
phase, it cannot clearly distinguish it from the localized phase. This
inability of the network to produce results similar to multi-class
classification is expected; as a matter of fact, the close connection
between the phase diagram obtained by the binary classifer to the
actual one is quite remarkable.

\section{Conclusion}\label{sec:level5}
In this work, we explore the ability of artificial neural networks to
extract information about various phases from single-particle
states. We build a multi-layer perceptron network and employ it to
classify the delocalized, multifractal and localized phases of the
quasiperiodic long-range Harper model. Our neural network produces
phase diagrams that align with theoretical predictions after being
trained using the data associated with the three phases. We establish
that the machine successfully learns each phase's property from the
eigenstates. The machine has high confidence throughout all the
phases, which is an indication of good feature extraction.

We also build a binary classifier with a single hidden layer having
probability densities of the eigenstates of the AAH model as inputs to
identify the delocalized and localized regimes. When we test this
neural network with the AAH model, we find that the resulting phase
diagram agrees very closely with the theoretically known phase
diagram. Even though the network is not trained with multifractal
states, the phase diagram obtained indicates the presence of
multifractal states, in a manner very similar to the actual phase
diagram. Subsequently, the network is applied to the LRH model for
detecting an unlearned phase. The neural network shows confusion,
whenever multifractal states are involved; however the phase diagram
obtained is remarkably close to the actual one. 
In this work, we can obtain the phase diagrams with reasonable accuracy by feeding a single disorder realization to the trained neural network. Thus, the neural network
method provides an alternative to the known conventional methods. Also, in 
recent times several works have explored the implementation of machine learning techniques 
in the context of many-body quantum systems~\cite{Carleo2018Constructing, Gardas2018Quantum, Carrasquilla2021How, Schmitt2020Quantum, Carleo2017Solving,doi:10.1126/science.aag2302, PhysRevB.95.245134}. A possible extension of our current 
work would be to identify and distinguish various phases in the many-body interacting system
 once the neural network learns the characteristics of the many-body
  wavefunctions in the different phases.

The phase classification problems in the literature have primarily
focused on binary classification. In this context, investigating
multiple phase transitions and training partially blind networks to
recognize unknown phases remains unexplored. The value of this
multi-neuron output strategy will be much more significant when
dealing with novel phases for which acceptable order parameters are
not known in advance.

\begin{acknowledgments}
A.A. is grateful to the Council of Scientific and Industrial Research
(CSIR), India, for her PhD fellowship. A.S. acknowledges financial
support from SERB via the grant (File Number: CRG/2019/003447) and DST
via the DST-INSPIRE Faculty Award [DST/INSPIRE/04/2014/002461]. All
neural networks used in this work were implemented in Python using
TENSORFLOW~\cite{45166}.
\end{acknowledgments}

\appendix
\section{3-D Anderson model}
  In this section, we employ the binary classifier to
  plot the phase diagram of the $3-$D Anderson
  model~\cite{PhysRev.109.1492}. In particular, we train the neural
  network (shown in Fig.~\ref{fig3}) by utilizing the eigenstate
  probabilities of the $1-$D AAH model~(Eq.~\ref{eq2}) with $N=512$
  sites. Since the dimension of the neural network's input layer is
  $512$, we consider a cubic lattice with $8^3=512$ sites to obtain
  the network's prediction.  The Hamiltonian of the $3-$D Anderson
  model is:
\begin{equation}
\hat{H}=J\sum_{\left\langle i,j \right\rangle}(\hat{c}_i^{\dagger}\hat{c}_{j}+\text{H.c.})+\Delta\sum_{i=1}
 \epsilon_i \hat{c}_i^{\dagger}\hat{c}_{i},
\end{equation}
 where the on-site energies $\epsilon_i=\left[-1/2,1/2 \right]$ are
 drawn from a uniform random distribution and $\Delta$ is the disorder
 parameter. Here, $J$ is set as unity.

\begin{figure}[b]
\centering
\stackunder{\hspace{-5.8cm}(a)}{\includegraphics[width=6cm]{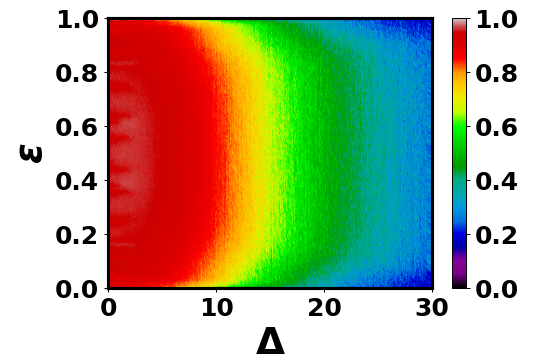}}\vspace{-0.5cm}
\stackunder{\hspace{-5.8cm}(b)}{\includegraphics[width=6cm]{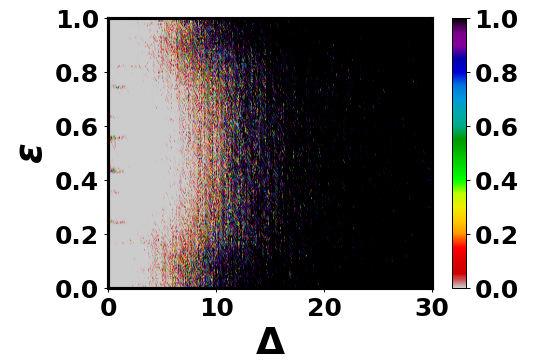}}
\caption{\label{fig9}(a)~Fractal dimension $D_2$ (whose value is
  represented by a color according to the code shown) as a function
  of disorder strength $\Delta$ and increasing fractional eigenstate
  index $i/N$ starting from the ground state for the 3-D Anderson model. 
  Here, averaging has been performed over $100$ disorder realizations. (b)~Classification
  of the eigenstates using the network
  trained on the central eigenstates of the 
  AAH model of system size $N=512$. The probability of the state being
  localized is given by the single output $P$, whose value is
  represented by a colour according to the code shown. Here, the system size of the $3-$D Anderson model is $8^3= 512$.}
\end{figure}

  In Fig~\ref{fig9}(a), we plot the phase diagram of the
  $3-$D Anderson model, with increasing strength of disorder $\Delta$
  and the color denoting the multifractal dimension $D_2$. 
  Above the critical disorder strength $\Delta \approx
  16.5$~\cite{MOTT19681}, all the eigenstates are exponentially localized
  with $D_2$ close to zero. At sub-critical disorder strengths, 
  localized and extended states are observed, separated by 
  some critical energy, dubbed the mobility edge~\cite{Anderson}.
  We employ the binary classifier by training it on all the eigenstates of the AAH model at
  the disorder strengths $\lambda=0.5$ (delocalized phase) and
  $\lambda=4$ (localized). The resulting network is then used to
  predict the phase diagram of the $3-D$ Anderson model, as shown in
  Fig.~\ref{fig9}(b). We observe that the network precisely identifies
  the delocalized eigenstates with $P\approx0$, the mobility edges with $0<P<1$, separating the 
  delocalized and localized eigenstates as well as the critical disorder strength 
   $\Delta \approx 16.5$ beyond which all the eigenstates are localized with
  $P$ close to unity. Thus, the network predicts the phase diagram of
  a system subjected to random disorder with reasonable accuracy
  despite the fact that it was trained on the eigenstates of a system
  with a quasiperiodic disorder and whose geometry is set in a
  different dimension.
 
\begin{figure}
\centering
\stackunder{\hspace{-3.8cm}(a)}{\includegraphics[width=4.2cm]{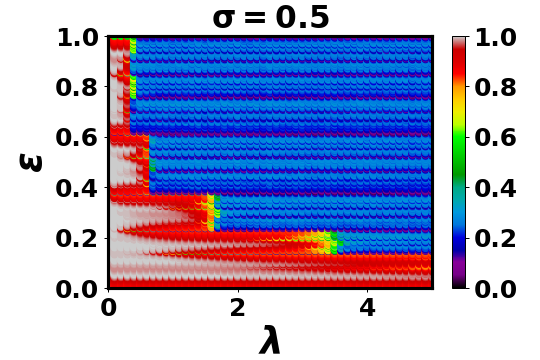}}
\stackunder{\hspace{-3.8cm}(b)}{\includegraphics[width=4.2cm]{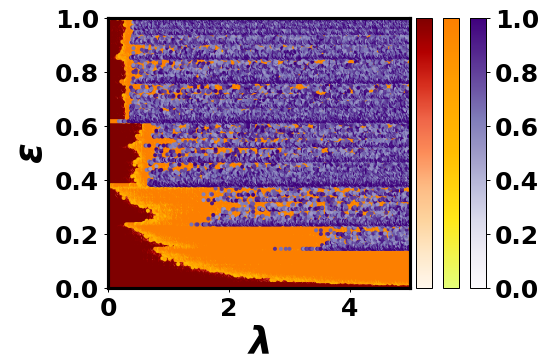}}
\vspace{-0.4cm}

\stackunder{\hspace{-3.8cm}(c)}{\includegraphics[width=4.2cm]{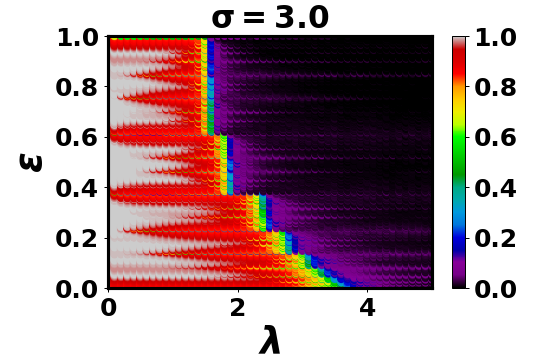}}
\stackunder{\hspace{-3.8cm}(d)}{\includegraphics[width=4.2cm]{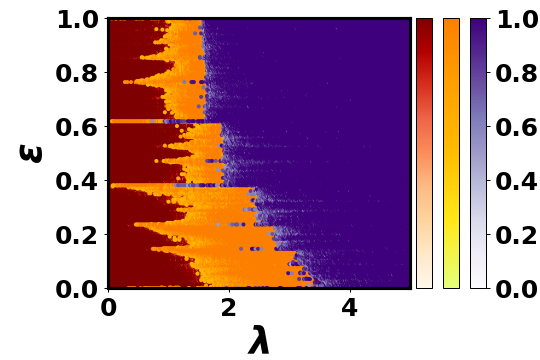}}
\vspace{-0.5cm}

\stackunder{\hspace{-3.8cm}(e)}{\includegraphics[width=4.2cm]{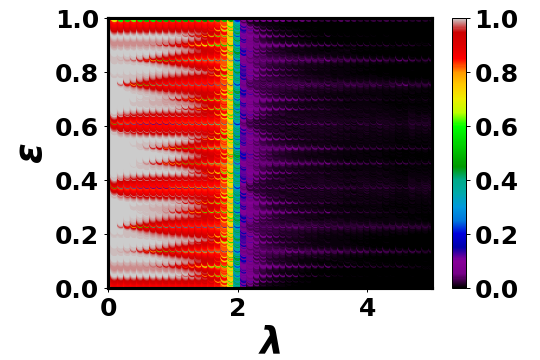}}
\stackunder{\hspace{-3.8cm}(f)}{\includegraphics[width=4.2cm]{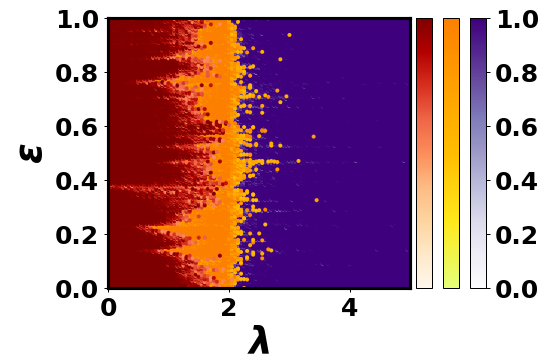}}
\caption{\label{fig10} Fractal dimension $D_6$ (whose value is
  represented by a colour according to the code shown) as a function
  of disorder strength $\lambda$ and for increasing fractional
  eigenstate index $i/N$ starting from the ground state for the
  long-range hopping parameter $\sigma$ equal to (a)~$0.5$, (c)~$3.0$ and (e)~AAH model. 
  Here, averaging has been performed over $100$ disorder
  realizations. Classification of the single disorder realization of
  the LRH model using the network trained on the eigenstates of the
  LRH model for hopping parameter $\sigma$ equal to (b)~$0.5$, (d)~$3.0$ and (f)~AAH model. The colour of each point
  $(\sigma,\lambda)$ represents the confidence of the network for the
  delocalized (red), multifractal (orange), and localized (purple)
  phases. Here, the system size is $N=510$ in all cases.}
\end{figure}
 
\section{Higher order fractal dimension}
In this section, we utilize higher-order fractal dimension 
$D_6$ (see Eq.~\ref{eq6}) to label the eigenstates as delocalized, localized and
multifractal in order to train the neural networks. For each eigenstate, the fractal dimension $D_6$ is calculated
and labelled as follows: 
\begin{align}
\nonumber D_6 <0.2 & \qquad \text{Localized,}\\
\nonumber 0.2 \leq D_6 \leq 0.8 & \qquad \text{Multifractal,}\\
D_6 > 0.8 & \qquad \text{Delocalized}.   
\end{align}
In Figs.~\ref{fig10}(a) and \ref{fig10}(c)
we have calculated $D_6$ as a function of disorder strength $\lambda$
for all the single particle eigenstates for the long-range hopping
parameter $\sigma=0.5$ and $3.0$. The corresponding phase diagrams 
plotted using the multi-class classification neural network trained using the
eigenstates of the LRH model where the class is assigned using $D_6$
are shown in Figs.~\ref{fig10}(b) and \ref{fig10}(d). We next compute the multifractal dimension $D_6$ as a function of
disorder strength $\lambda$ for all the single particle eigenstates of
the AAH model (see Fig.~\ref{fig10}(e)). The corresponding phase diagram
obtained using the neural network trained on the
eigenstates of the LRH model is shown in Fig.~\ref{fig10}(f). We observe 
that even when the network is trained on the eigenstates labelled using higher
moments, such as $D_6$, it classifies the various phases with
reasonable accuracy, as well as precisely predicts the location of the
transitions/steps. This implies that any quantity which reflects the 
localization properties of the system can be used for labelling data in supervised machine learning.

\bibliography{ml}
\end{document}